\documentclass[12pt]{article}
\usepackage{graphicx}
\def \s{\sqrt{2}}
\def \bea{\begin{eqnarray}}
\def \eea{\end{eqnarray}}
\def \eeq{\end{equation}}
\def \beq{\begin{equation}}
\def\tl{\tilde\lambda}
\def\lts{\lambda_t^{(s)}}

\def\lusstar{\lambda_u^{(s)*}}
\def\lus{\lambda_u^{(s)}}
\def\lud{\lambda_u^{(d)}}
\def\ltd{\lambda_t^{(d)}}
\def\b{\cal B}

\begin{document}
\rightline{TECHNION-PH-2009-34}
\rightline{arXiv:1001.0702}
\rightline{January 2010} 
\bigskip
\centerline{\bf CP ASYMMETRIES IN $B\to K\pi, K^*\pi, \rho K$ DECAYS} 
\bigskip 
 
\centerline{Michael Gronau} 
\centerline{\it Physics Department, Technion -- Israel Institute of Technology} 
\centerline{\it 32000 Haifa, Israel} 
\medskip 
 
\centerline{Dan Pirjol} 
\centerline{\it Department of Particle Physics, National Institute for Physics and Engineering} 
\centerline{\it  077125 Bucharest, Romania} 
\medskip

\centerline{Jure Zupan} 
\centerline{\it Faculty of Mathematics and Physics, University of Ljubljana} 
\centerline{\it Jadranska 19, 1000 Ljubljana, Slovenia} 
\centerline{and}
\centerline{\it Josef Stefan Institute, Jamova 39, 1000 Ljubljana, Slovenia}
 
\begin{quote}
We show that ratios of tree and penguin amplitudes in $B\to K^*\pi$ and 
$B\to \rho K$ are two to three times larger than in $B\to K\pi$.  
This allows for considerably larger CP asymmetries  in the former processes 
than the $10\%$ asymmetry measured in $B^0\to K^+\pi^-$. 
We study 
isospin sum rules for rate asymmetries in $B\to K\pi, K^*\pi, \rho K$, estimating  
small violation from interference of tree and electroweak penguin 
amplitudes. The breaking of the $K\pi$ asymmetry sum rule is  
estimated to be one to two percent and negative. 
Violation  of $K^*\pi$ and $\rho K$ sum rules can be estimated from $B\to \rho\pi$  
amplitudes using  flavor SU(3), while breaking of a sum rule combining 
$K^*\pi$ and $\rho K$ asymmetries can be
measured directly in a Dalitz analysis of $B^0\to K^+\pi^-\pi^0$. 
The three sum rules can be tested using complete sets of data taken at 
$e^+e^-$ $B$ factories and in experiments at the 
LHCb and at a future Super Flavor Factory, providing precision searches for 
new $\Delta S = \Delta I=1$ operators in the low energy effective Hamiltonian.
\end{quote}
\bigskip

{\bf I. INTRODUCTION}

\medskip
The Cabibbo-Kobayashi-Maskawa (CKM) framework for flavor physics and CP violation
has been confirmed successfully in numerous experiments involving a variety of $B$ 
meson decays~\cite{PDG,Charles:2004jd,Bona:2009ze,HFAG}. 
The measured 
oscillating CP asymmetry in $B^0\to J/\psi K_{S,L}$ has a theoretically precise interpretation 
in terms of the CKM phase $\beta$, at a level of $10^{-3}$, including a small $b\to u\bar us$ 
amplitude whose calculation involves a perturbative term~\cite{Boos:2004xp} and long distance rescattering corrections~\cite{Gronau:2008cc}. This asymmetry provided an accurate value for 
$\beta$ with an experimental precision of $1^\circ$. Using isospin symmetry~\cite{Gronau:1990ka,Snyder:1993mx},
time-dependent CP symmetries measured in $B^0\to \pi^+\pi^-, \rho^+\rho^-, \rho^\pm\pi^\mp$
 led to a determination of the phase $\alpha$ with an error of $4^\circ$. The values obtained for the 
 two CP-violating phases are in good agreement with CP-conserving CKM constraints. 
Other asymmetries probing the phase $\beta$ were measured 
in loop-dominated processes from $b\to q\bar qs$ ($q=u, d, s$) including $B^0\to \pi^0 K_S, 
\eta' K_S, \phi K_S, \rho^0 K_S$. These processes are sensitive to corrections from
flavor changing $\Delta S=1$ operators appearing in extensions of the low energy 
theory~\cite{Gronau:1996rv,Grossman:1996ke}. 
Uncertainties in calculating hadronic amplitudes by applying QCD 
dynamics~\cite{Beneke:2005pu,Cheng:2005bg}
prohibit  distinguishing between New Physics and hadronic effects on these asymmetries.

Direct CP asymmetries have been observed in several charmless $\Delta S=1$ $B$ decays, 
with a high confidence level above $8\sigma$ in $B^0\to K^+\pi^-$ and at a lower confidence
 level  of $3$-$4\sigma$ in  
$B^+\to \eta K^+, \rho^0K^+$ and $B^0\to \eta K^{*0}$~\cite{HFAG}. These observations are 
interesting by themselves but do not provide clean tests for the CKM framework. An 
interpretation of direct asymmetries in terms of fundamental CP phases requires calculating
nonleptonic decay amplitudes and strong phases.
Remarkable progress has been achieved in the past ten years in the theory of hadronic $B$ 
decays, starting with work advocating QCD factorization~\cite{Beneke:1999br} and 
perturbative QCD~\cite{Keum:2000ph}. While great progress has been made recently in 
calculating higher order terms in 
$\alpha_s$~\cite{Bell:2007tv,Beneke:2009ek}, $1/m_b$ corrections remain a serious difficulty.
Thus, nonfactorizable hadronic matrix elements of charming penguin 
operators~\cite{Ciuchini:1997hb,Bauer:2004tj} and of color-suppressed tree amplitudes  
lead to difficulties in calculating the decay rate for 
$B^0\to \pi^0\pi^0$~\cite{Beneke:1999br,Beneke:2009ek,Bell:2009fm}.  
This  example and an intrinsic failure to account for individual strong phases (as discussed briefly 
in Sec. II for one example), seem to originate in incalculable long distance final state 
rescattering effects. 
 
 Whereas individual direct CP asymmetries in hadronic $B$ decays cannot be calculated 
 reliably, there are certain classes 
 of decays in which asymmetries can be related to each other within the CKM framework 
 in a model independent way using symmetry arguments. Isospin symmetry, which is expected 
 to hold within a couple of percent, has been shown to imply an approximate sum rule for 
 asymmetries in the four $B\to K\pi$ decay processes~\cite{Gronau:2005kz},
 \beq\label{SRasym}
A_{CP}(K^+\pi^-) + A_{CP}(K^0\pi^+) \approx A_{CP}(K^+\pi^0) + A_{CP}(K^0\pi^0)~.
\eeq
A violation of this sum rule would be evidence for a new $\Delta S = \Delta I=1$ term
in the low enery effective Hamiltonian. A similar sum rule may hold approximately 
in the CKM framework for $B\to K^*\pi$ and $B\to\rho K$ decays, where individual CP 
asymmetries may be larger than in $B\to K\pi$.  
 
In this paper we will study  asymmetries in penguin-dominated $B\to K\pi, K^*\pi$ 
and $B\to \rho K$ and will compare their three isospin sum rules. These sum rules involve 
corrections from interference of subleading tree and electroweak penguin (EWP) amplitudes
which are related to each other in a flavor SU(3) limit. Our aim will be  to estimate these second 
order corrections in a model-independent way, or to propose methods in which the corrections can
be measured elsewhere.
Our main results are summarized in Eqs.~(65), (72), (81)-(83).
In order to present a self-contained  analysis for the three sum rules in 
$B\to K\pi, B \to K^*\pi$ and $B\to \rho K$ we will introduce and study 
in some detail amplitudes and ratios of amplitudes occurring  in the three classes of 
decay processes.  

Section II gives $B\to K\pi$ decay amplitudes in terms of their diagramatic contributions, 
discussing the role of these contributions in $B\to K\pi$ asymmetries. Sec. III 
generalizes this description to $B\to K^*\pi$ and $B\to\rho K$ decays. In Sec. IV we use 
broken  flavor SU(3) to calculate ratios of tree-to-penguin amplitudes in $B\to K\pi, 
K^*\pi, \rho K$, providing estimates for maximal possible CP asymmetries 
in these three classes of processes. Sec. V studies experimental tests for the broken SU(3) 
symmetry assumption. SU(3) relations between subleading tree and EWP amplitudes are 
discussed in Sec. VI, and are used in Sec. VII for estimating deviations from exact sum rules among 
asymmetries for $B\to K\pi, K^*\pi$ and $B\to \rho K$ decays. Sec. VIII concludes.
\bigskip

{\bf II. AMPLITUDES AND ASYMMETRIES IN $B\to K\pi$}

\medskip
Decay amplitudes for $B\to K\pi$ may be described generally in a model-independent way
using  topological graphical contributions~\cite{Gronau:1994rj},
\bea\label{+-}
-A(K^+\pi^-) & = & \lambda_t^{(s)}(P_{tc} + \frac{2}{3}P^C_{EW})
+ \lambda_u^{(s)}(P_{uc} + T) ~,\\
\label{0+}
A(K^0\pi^+) & = & \lambda_t^{(s)}(P_{tc} - \frac{1}{3} P^C_{EW})
+ \lambda_u^{(s)}(P_{uc} + A)~,\\
\label{+0}
-\s A(K^+\pi^0) & = & \lambda_t^{(s)}(P_{tc} + P_{EW} + \frac{2}{3}P^C_{EW})
+ \lambda_u^{(s)}(P_{uc} + T + C + A) ~,\\
\label{00}
\s A(K^0\pi^0) & = & \lambda_t^{(s)}(P_{tc} - P_{EW} - \frac{1}{3}P^C_{EW})
+  \lambda_u^{(s)}(P_{uc} - C)~,
\eea
where $\lambda_q^{(q')}\equiv V^*_{qb}V_{qq'}~(q=u, t; q'=d, s)$ are 
Cabibbo-Kobayashi-Maskawa (CKM) factors with a small ratio
$|\lambda_u^{(s)}|/|\lambda_t^{(s)}| \sim 0.02$. 
Each of the seven amplitudes, $P_{tc}, P_{EW}, P^C_{EW}, T, C,$ $P_{uc}, A$ involves 
an unknown strong phase.

The common penguin amplitude $P' \equiv \lambda_t^{(s)}P_{tc}$ [or the linear combination 
$\lambda_t^{(s)}(P_{tc} - P^C_{EW}/3)$] with weak phase 
${\rm arg}(\lambda_t^{(s)}) =\pi$, dominates all four $B\to K\pi$ amplitudes. This dominance 
is tested  by simple approximate ratios,
\beq\label{ratios}
{\cal B}(K^+\pi^-):{\cal B}(K^0\pi^+)/r_{\tau}:{\cal B}(K^+\pi^0)/r_{\tau}:{\cal B}(K^0\pi^0) \simeq
1:1:\frac12:\frac12~,
\eeq
The four branching ratios given in Table I, and the $B^+$ to $B^0$ lifetime ratio, 
$r_{\tau}\equiv \tau_+/\tau_0 =1.071 \pm 0.009$~\cite{HFAG}, imply experimental ratios:
 \beq\label{ratios-numbers}
(0.899\pm 0.048) : 1 : (0.558 \pm 0.035) : (0.454\pm 0.034)~.
\eeq
Thus Eq.~(\ref{ratios}) holds reasonably well with several percent errors leaving little space for
non-penguin amplitudes.
%
\begin{table}[h]
\caption{Branching fractions and CP asymmetries for $B\to K\pi, K^*\pi, \rho K$~\cite{HFAG}.
\label{tab:kkstrho1}}
\begin{center}
\begin{tabular}{c c c} \hline \hline
Mode & $\b$~$(10^{-6})$ & $A_{CP}$  \\ \hline\hline
$B^0 \to K^+\pi^-$ & $19.4 \pm 0.6$ & $-0.098^{+0.012}_{-0.011}$ \\
$B^+ \to K^0 \pi^+$ & $23.1 \pm 1.0$  & $0.009 \pm 0.025$\\
$B^+ \to K^+\pi^0$ & $12.9 \pm 0.6$ & $0.050 \pm 0.025$ \\
$B^0 \to K^0\pi^0$ &  $9.8 \pm 0.6$  & $-0.01 \pm 0.10$ \\  \hline
$B^0\to K^{*+}\pi^-$ & $8.6^{+0.9}_{-1.0}$ & $-0.18 \pm 0.08$\\
$B^+\to K^{*0}\pi^+$ & $9.9^{+0.8}_{-0.9}$ & $-0.038\pm 0.042$\\
$B^+\to K^{*+}\pi^0$ & $6.9 \pm 2.3$  & $0.04\pm 0.29$\\ 
$B^0\to K^{*0}\pi^0$ & $2.4\pm 0.7$ & $-0.15\pm 0.12$\\ \hline
$B^0\to \rho^-K^+$ & $8.6^{+0.9}_{-1.1}$ & $0.15\pm 0.06$\\
$B^+\to\rho^+K^0$ & $8.0^{+1.5}_{-1.4}$ &  $-0.12 \pm 0.17$\\
$B^+\to \rho^0K^+$ & $3.81^{+0.48}_{-0.46}$ & $0.37 \pm 0.11$\\
$B^0\to\rho^0K^0$ & $4.7\pm 0.7$ & $0.06\pm 0.20$\\
\hline \hline
\end{tabular}
\end{center}
\end{table}

Let us now discuss the subdominant terms occurring in Eqs.~(\ref{+-})-(\ref{00}), their relative
magnitudes with respect to $P_{tc}$ or $P'$, and their effects
on CP asymmetries in $B\to K\pi$ decays.
\begin{itemize}
\item The largest subdominant term with weak phase ${\rm arg}(\lambda_u^{(s)}) \equiv \gamma$ 
is expected to be the color-favored tree amplitude $T' \equiv \lambda_u^{(s)}\, T$. An interference between $P'$ and $T'$ dominates the asymmetry in $B^0\to K^+\pi^-$ for which a 
negative value has been measured at a level of $10\%$. We will show in the next section that 
$T'$ is about an order of magnitude times smaller that $P'$. 
\item The difference between the asymmetry in $B^0\to K^+\pi^-$  and the preferably positive 
asymmetry measured in 
$B^+\to K^+\pi^0$, obtaining contributions from  interference of $P'$ with both $T'$ and and 
with the formally color-suppressed $C'\equiv \lambda_u^{(s)}\,C$, has 
shown~\cite{Gronau:2006ha,Fleischer:2007mq} 
that the ratio $|C|/|T|$ is not much smaller than one as naively anticipated, and that the relative 
strong phase ${\rm Arg}(CT^*)$ is sizable and negative.  (We use a convention 
in which strong phase differences lie between $-\pi$ and $\pi$.)
We note that ${\rm Arg}(CT^*) \simeq 0$ is predicted at leading order in $1/m_b$ by QCD 
factorization~\cite{Beneke:2001ev}
and in a Soft Collinear Effective Theory approach~\cite{Bauer:2004tj}.
The large strong phase may be interpreted as either large $1/m_b$ corrections 
or sizable nonfractorizable contributions to $C$ from rescattering through 
color-favored  intermediate states. 
\item EWP terms $P_{EW}, P_{EW}^C$, which are higher order in the electroweak coupling,
are an order of magnitude times smaller than $P_{tc}$. A quantitative discussion of these contributions relating 
them to penguin and tree amplitudes will be presented in Section VI. Effects of EWP terms on CP asymmetries through their interference with tree amplitudes are of second order because these 
two kinds of amplitudes are an order of magnitude smaller than $P_{tc}$. Interference of EWP
amplitudes with $P_{tc}$ leads to no CP asymmetry as these two contributions carry the same
weak phase.
\item The very small asymmetry measured in $B^+\to K^0\pi^+$, consistent with zero,
indicates that both $P_{uc}$ and the annihilation amplitude $A$ are much smaller  than $T$. 
A large enhancement by rescatterring of these amplitudes, which are expected to be intrinsically very 
small~\cite{Blok:1997yj},  would have created a sizable strong phase in these amplitudes relative to 
$P_{tc}$, thereby leading to a non-nlegligible CP asymmetry in $B^+\to K^0\pi^+$. 
\end{itemize}

\bigskip
{\bf III. AMPLITUDES IN $B\to K^*\pi$ and $B\to \rho K$}

\medskip
 Amplitudes for the eight processes $B\to K^*\pi$ and $B\to \rho K$  with final charges as in 
$B\to K\pi$ may be decomposed into expressions similar to Eqs.~(\ref{+-})--(\ref{00}), where 
$P, T, C, A$ are now replaced by $P_P, T_P, C_V, A_P$ in $B\to K^*\pi$ and by 
$P_V, T_V, C_P, A_V$ in $B\to \rho K$~\cite{Dighe:1995gq}. The suffix $M= P, V$ 
denotes whether the spectator quark is included in a pseudoscalar ($P$)
or vector meson  ($V$). Electroweak contributions, $P'_{EW,M} \equiv \lambda_t^{(s)}\,P_{EW,M}$ 
and $P'^C_{EW,M} \equiv \lambda_t^{(s)}\,P^C_{EW,M}$,  are introduced as in $B\to K\pi$
through the substitution~\cite{Gronau:1994rj}
\beq
T'_M \to T'_M + P'^C_{EW,M}~,~~~C'_M \to C'_M  + P'_{EW,M}~,~~
P'_M \to P'_M -\frac13P'^C_{EW,M}~.
\eeq

Thus, $B\to K^*\pi$ amplitudes analogous to Eqs.~(\ref{+-})-(\ref{00}) are given by:
\bea\label{K*+-}
-A(K^{*+}\pi^-) & = & \lambda_t^{(s)}(P_{tc,P} + \frac{2}{3}P^C_{EW,P})
+ \lambda_u^{(s)}(P_{uc,P} + T_P) ~,\\
\label{K*0+}
A(K^{*0}\pi^+) & = & \lambda_t^{(s)}(P_{tc,P} - \frac{1}{3} P^C_{EW,P})
+ \lambda_u^{(s)}(P_{uc,P} + A_P)~,\\
\label{K*+0}
-\s A(K^{*+}\pi^0) & = & \lambda_t^{(s)}(P_{tc,P} + P_{EW,V} + \frac{2}{3}P^C_{EW,P})
+ \lambda_u^{(s)}(P_{uc,P} + T_P + C_V + A_P) ~,\\
\label{K*00}
\s A(K^{*0}\pi^0) & = & \lambda_t^{(s)}(P_{tc,P} - P_{EW,V} - \frac{1}{3}P^C_{EW,P})
+  \lambda_u^{(s)}(P_{uc,P} - C_V)~.
\eea
Corresponding amplitudes for $B\to K\rho$ are obtained by interchanging subscripts 
$P\leftrightarrow V$.

Penguin dominance in $B\to K^*\pi$ and $B\to \rho K$ decays leads to approximate ratios 
of branching ratios in these processes which are similar to the ratios $1:1:1/2:1/2$ in $B\to K\pi$. 
Using branching ratios in Table I, involving experimental errors  considerably larger 
than in $B\to K\pi$, one finds, 
\bea\label{ratiosK*pi}
&&{\cal B}(K^{*+}\pi^-):{\cal B}(K^{*0}\pi^+)/r_{\tau}:{\cal B}(K^{*+}\pi^0)/r_{\tau}:{\cal B}(K^{*0}\pi^0)
\nonumber\\
&&= (0.93 \pm 0.13) : 1 : (0.70 \pm 0.24) : (0.26 \pm 0. 08)~,
\eea
\bea\label{ratiosrhoK}
&&{\cal B}(\rho^-K^+):{\cal B}(\rho^+K^0)/r_{\tau}:{\cal B}(\rho^0K^+)/r_{\tau}:{\cal B}(\rho^0K^0)
\nonumber\\
&&= (1.15 \pm 0.25) : 1 : (0.48 \pm 0.10) : (0.63 \pm 0.15)~.
 \eea
The value of ${\cal B}(K^{*0}\pi^0)$ in (\ref{ratiosK*pi}) seems somewhat 
small~\cite{Gronau:2005ax} because it involves small values with large errors in old CLEO 
and Belle measurements~\cite{Jessop:2000bv,Chang:2004um}.
Using the considerably more precise Babar measurement~\cite{Aubert:2007bs}, 
${\cal B}(K^{*0}\pi^0) = (3.6 \pm 0.7 \pm 0.4)\times 10^{-6}$, the last numerical factor
in (\ref{ratiosK*pi}) becomes $0.39 \pm 0.09$ consistent within error with $1/2$. 
We note that large experimental errors in branching ratio measurements
leave ample space for subdominant non-penguin amplitudes in these processes. 

Magnitudes of dominant penguin amplitudes and of subdominant amplitudes, and certain 
relative strong phases between them, have been 
studied numerically in Ref.~\cite{Chiang:2003pm}, applying a global flavor SU(3) fit 
to data of these processes and corresponding $\Delta S=0$ $B\to VP$ decays. 
Recently $\chi^2$ fits were performed for $B\to K^*\pi$ and $B\to \rho K$ data, concluding that 
current experimental errors are too large for showing an inconsistency with the 
CKM framework~\cite{Chiang:2009hd}. 
The purpose of the next section is limited to estimating ratios of tree and penguin amplitudes determining the potentially largest CP asymmetries in $B\to K^*\pi$ and $B\to \rho K$.

\bigskip
{\bf IV. RATIOS OF TREE-TO-PENGUIN AMPLITUDE IN $B\to K\pi, K^*\pi, \rho K$}

\medskip
The ratios $|T'/P'|, |T'_P/P'_P|$ and $|T'_V/P'_V|$  determine the 
maximal potential asymmetries in $B^0\to K^+\pi^-$, $B^0\to K^{*+}\pi^-$
and $B^0\to \rho^-K^+$, given roughly by $ 2|T'/P'|\sin\gamma$, $2|T'_P/P'_P|\sin\gamma$
and $2|T'_V/P'_V|\sin\gamma$. Values of these three ratios may be estimated by relating 
within broken flavor SU(3) the amplitudes for  these processes to those for $B^0\to \pi^+\pi^-, 
B^0\to \rho^+\pi^-$ and $B^0\to \rho^-\pi^+$. 
One way of using flavor SU(3) is by applying group theory for decomposing $\Delta S=1$ and 
$\Delta S=0$ decay amplitudes into five SU(3)  reduced matrix elements in $B\to PP$  
and ten reduced matrix elements in $B\to VP$~\cite{Zeppenfeld:1980ex}. 
A more powerful SU(3) method is based on a diagramatic  
approach~\cite{Gronau:1994rj,Dighe:1995gq} in which a certain hierarchy between 
amplitudes can be shown to exist~\cite{Bauer:2004ck} and SU(3) breaking factors in  
tree amplitudes may be motivated by a factorization assumption.
In the diagramatic approach, where certain combinations of terms correspond to 
SU(3) reduced matrix elements,  tree amplitudes defines as~\cite{Dighe:1995gq},
\beq
T'_P = T_P + P_{uc,P}~,~~~~~~~~~~~~T'_V = T_V + P_{uc,V}~,
\eeq
are well-defined, and are renormalization-group scale and scheme-independent.

We keep SU(3) invariant penguin amplitudes as a default because factorization is not 
expected to hold for these contributions~\cite{Ciuchini:1997hb,Bauer:2004tj}. 
Tests of these assumptions will be presented in the next section. The effect of possible
SU(3) breaking in penguin amplitudes, at an expected level of $20\%$, can be easily included 
in our estimates for the ratios of tree and penguin amplitudes.

\medskip
\begin{table}[h]
\caption{Amplitudes, branching fractions and asymmetries for certain $B\to PP, VP$~\cite{HFAG}.
\label{tab:kkstrho2}}
\begin{center}
\begin{tabular}{c c c c } \hline \hline
Mode & Amplitude & $\b$~$(10^{-6})$ & $A_{CP}$ \\ \hline\hline
$B^+ \to K^0 \pi^+$ & $P'$  & Table I & Table I\\
$B^0\to K^+\pi^-$ & $-(P' + T')$ & Table I & Table I \\
$B^0 \to \pi^+\pi^-$ & $\tl P' - \tl^{-1}T'\left (\frac{f_\pi}{f_K}\right)$ & $5.16 \pm 0.22$
& $0.38\pm 0.06$ \\ \hline
$B^+ \to K^{*0}\pi^+$ & $P'_P$ & Table I & Table I \\
$B^0\to K^{*+}\pi^-$ & $-(P'_P + T'_P)$ & Table I & Table I \\
$B^0 \to \rho^+\pi^-$ &  $ \tl P'_P - \tl^{-1}T'_P\left(\frac{f_\rho}{f_{K^*}}\right)$  & 
$15.7\pm 1.8^a$
& $0.11 \pm 0.06$ \\ \hline
$B^+ \to \rho^+K^0$ & $P'_V$ & Table I & Table I \\
$B^0\to \rho^-K^+$ & $-(P'_V + T'_V)$ & Table I & Table I \\
$B^0 \to \rho^-\pi^+$ &  $\tl P'_V - \tl^{-1}T'_V\left(\frac{f_\pi}{f_K}\right)$  & $7.3\pm 1.2^c$
& $ -0.18 \pm 0.12$ \\ 
\hline \hline
\end{tabular}
\end{center}
\leftline{$^{a,c}$ We use ${\cal B}(\rho^\pm\pi^\mp)= 
\frac 12{\cal B}(1 \pm A_{CP}^{\rho\pi}C \pm \Delta C)$,
${\cal B}\equiv {\cal B}(\rho^+\pi^-) + {\cal B}(\rho^-\pi^+)$.}
\end{table}

Table II lists dominant terms in amplitudes (in the 
$c$-convention~\cite{Gronau:2002gj}), branching ratios and asymmetries for $\Delta S=1$ penguin-dominated  amplitudes, and for $\Delta S=0$ processes involving the three pairs $(P', T'), (P'_P, T'_P), (P'_V, T'_V)$.
We define $\tl \equiv \lambda/(1-\lambda^2/2)=0.232$, where $\lambda$ is the Wolfenstein 
parameter~\cite{Wolfenstein:1983yz}, and use the following ratios of meson decay constants to represent SU(3) breaking factors in color-favored tree amplitudes,
 $f_\pi/f_K = 0.84$,  $f_\rho/f_{K^*} = 0.96$~\cite{PDG}. 
 Other SU(3) breaking factors in tree amplitudes involving ratios of kernels and ratios of their overlaps with meson wave functions (given roughly by ratios of a given form factor at  slightly different values of $q^2$) will be neglected. 
 
 In order to give simple estimates for $|T'_{(V,P)}|/|P'_{(V,P)}|$ we first assume that 
\beq\label{P'vsT'}
\tl^2 |P'_{(V,P)}| \ll |T'_{(V,P)}|~.
\eeq
Thus we estimate the three ratios $|T'_{(V,P)}|/|P'_{(V,P)}|$ using central values for 
branching ratios given in Tables I and II,
\bea\label{T'/P'}
\frac{|T'|}{|P'|} & \simeq & \tl\left (\frac{f_K}{f_\pi}\right ) 
\sqrt{\frac{r_\tau\b(\pi^+\pi^-)}{{\cal B}(K^0\pi^+)}} = 0.13~,\\
\label{T'_P/P'_P}
\frac{|T'_P|}{|P'_P|} & \simeq & \tl\left (\frac{f_{K^*}}{f_\rho}\right )
\sqrt{\frac{r_\tau\b(\rho^+\pi^-)}{{\cal B}(K^{*0}\pi^+)}} = 0.31~,\\
\label{T'_V/P'_V}
\frac{|T'_V|}{|P'_V|} & \simeq & \tl\left (\frac{f_K}{f_\pi}\right ) 
\sqrt{\frac{r_\tau\b(\rho^-\pi^+)}{{\cal B}(K^0\rho^+)}} = 0.27~.
\eea
We note that our assumption (\ref{P'vsT'}) is a good approximation for the last two cases
where indeed $|T'_{P,V}|/|P'_{P,V}| \gg \tl^2 = 0.05$, but may provide only a crude estimate 
for $|T'|/|P'|$.

One may relax the assumption (\ref{P'vsT'}) for $|T'|/|P'|$ by replacing  (\ref{T'/P'}) with a quadratic relation following from the amplitudes in Table II,
\beq
R \equiv \frac{r_\tau{\cal B}(\pi^+\pi^-)}{{\cal B}(K^0\pi^+)} = \tl^2 
+ \left (\frac{|T'|}{|P'|}\right )^2 \left (\tl\frac{f_K}{f_\pi}\right)^{-2}
+2\frac{|T'|}{|P'|}\frac{f_\pi}{f_K}\cos\delta\cos\gamma~,
\eeq 
where $\delta$ is the unknown strong phase difference between $P'$ and $T'$.
Denoting $z=\cos\delta\cos\gamma$ one solves for $|T'|/|P'|$,
\beq
\frac{|T'|}{|P'|} = \tl^2\frac{f_K}{f_\pi}\left (\sqrt{z^2 + (R-\tl^2)/\tl^2} - z\right )~,
\eeq
which is a monotonically decreasing function of $z$.
 Taking conservative bounds $-0.6 \le z \le 0.6$, based on a lower bound on 
 $\gamma$~\cite{Charles:2004jd} and on no restrictions on $\delta$, one obtains
 \beq\label{T/Pbound}
 0.09 \le \frac{|T'|}{|P'|} \le 0.16~,
 \eeq
 which describes a rather narrow range around the value in (\ref{T'/P'}).
 Similarly, including quadratic corrections in estimates of $|T'_P|/|P'_P|$ and $|T'_V|/|P'_V|$,
 one obtains the following bounds,
 \beq\label{TVP/PVPbounds}
 0.28 \le \frac{|T'_P|}{|P'_P|} \le 0.35~,~~~~~~0.23 \le \frac{|T'_V|}{|P'_V|} \le 0.31~.
  \eeq
We do not include errors in input branching ratios. The resulting uncertainties in the 
above two ratios from errors in branching ratios are somewhat smaller than the ones 
shown, which are caused by our conservative assumption of completely arbitrary strong  
phases  $\delta$. 

Our conclusion is that the ratios $|T'_P|/|P'_P|$ and $|T'_V|/|P'_V|$ of tree-to-penguin 
amplitudes in $B\to K^*\pi$ and $B\to \rho K$, respectively, are between two to three 
times larger than the corresponding ratio $|T'|/|P'|$ in $B\to K\pi$. The processes 
$B^0\to K^{*+}\pi^-$ and $B^+\to K^{*+}\pi^0$ or $B^0\to K^+\rho^-$ and $B^+\to K^+\rho^0$ 
involve interference of $P'_P$ and $T'_P$ or interference of $P'_V$ and $T'_V$. Thus, these 
processes may potentially involve asymmetries between two to three times larger than 
the $10\%$ asymmetry measured in $B^0\to K^+\pi^-$. 

Maximal CP asymmetries 
$A_{CP}^{\rm max} \simeq 2|T'_{(P,V)}|/|P'_{(P,V)}|\sin\gamma$ in $B^0\to K^+\pi^-, B^0\to K^{*+}\pi^-$ 
and $B^0\to \rho^-K^+$, are obtained for $\delta =90^\circ$ or $z=0$ corresponding to the central 
values of $|T'_{(P,V)}|/|P'_{(P,V)}|$ given in Eqs.~(\ref{T'/P'}), (\ref{T'_P/P'_P})  and (\ref{T'_V/P'_V}).
The actual values of these asymmetries depend, of course, on $\delta$. The measured asymmetry
in $B^0\to K^+\pi^-$ indicates $|T'/P'| \sim 0.10$ within the range (\ref{T/Pbound}) and
$\delta(K\pi) \sim 30^\circ$. The strong phases in $B^0\to K^{*+}\pi^-$ and $B^0\to \rho^-K^+$ 
may be different which would affect the asymmetries in these processes. 

The asymmetry in $B^+\to K^{*+}\pi^0$ ($B^+\to K^+\rho^0$) depends also on the 
interference of $P'_P$ and $C'_V$ ($P'_V$ and $C'_P$). Unlike $B^+\to K^+\pi^0$, where the
effect of $C'$ on the CP asymmetry is destructive relative to the effect of $T'$, these 
interference effects do not have to act destructively with respect to the interference between
$P'_P$ and $T'_P$ ($P'_V$ and $T'_V$) . Thus they may increase the asymmetry in 
$B^+\to K^{*+}\pi^0$ ($B^+\to K^+\rho^0$) relative to that in $B^0\to K^{*+}\pi^-$ 
($B^0\to K^+\rho^-$).

We note in passing that one may use the ranges of value in Eqs.~(\ref{T/Pbound}) 
and (\ref{TVP/PVPbounds}) to estimate ratios of penguin and tree amplitudes in $B^0\to \pi^+\pi^-$, $B^0\to \rho^+\pi^-$ and $B^0\to \rho^-\pi^+$~\cite{Gronau:2004ej,Gronau:2004tm}. Taking into 
account CKM and SU(3) breaking factors, one finds for these three ratios
\beq\label{P'/T'}
0.40 \le \tl^2\frac{f_K}{f_\pi}\frac{|P'|}{|T'|} \le 0.71~,~~~
0.16 \le \tl^2\frac{f_{K^*}}{f_\rho}\frac{|P'_P|}{|T'_P|} \le 0.20~,~~~
0.21 \le \tl^2\frac{f_K}{f_\pi}\frac{|P'_V|}{|T'_V|} \le 0.28~.
\eeq
Thus, whereas the penguin contribution in $B^0\to \pi^+\pi^-$ is only somewhat smaller
than that of the tree amplitude, penguin terms contribute only about $20\%$ to  the amplitudes
for $B^0\to\rho^+\pi^-$ and $B^0\to\rho^-\pi^+$ which are largely dominated by tree amplitudes.
We note that while calculations in Ref.~\cite{Beneke:2003zv} using QCD factorization obtain 
values for the ratios  (\ref{P'/T'}) which are about a factor two smaller than the above,
a more recent update agrees with the above ranges~\cite{Beneke:2006mk}.
\bigskip

{\bf V. TESTS OF FLAVOR SU(3)}

\medskip
In the previous section we assumed that penguin amplitudes are SU(3)-invariant, while 
SU(3) breaking factors involving ratios of meson decay constants were assumed for
tree amplitudes.  In this section we will present experimental tests for these assumptions based on 
measurements of decay rates and CP asymmetries in pairs of $\Delta S=1$ and $\Delta S=0$ 
decays which are related to each other by flavor SU(3).  

Consider the three pairs of $B^0$ decay processes listed in Table II. The structure of these  
amplitudes, involving for a given pair the same penguin and tree amplitudes with different 
CKM factors, leads to simple relations between CP rate differences defined by
\beq
\Delta(B \to f) \equiv \Gamma(\bar B \to \bar f) - \Gamma(B\to f) 
\equiv 2\bar\Gamma(B\to f)A_{CP}(B\to f)~,
\eeq
where $\bar\Gamma$ is a CP-averaged decay rate.
Denoting by ${\cal B}A_{CP}$ the product of a charge-averaged branching ratio and
a CP asymmetry for a given process,  one finds~\cite{Deshpande:1994ii,Gronau:1995qd}
\beq
\Delta(K^+\pi^-) = -\frac{f_K}{f_\pi}\Delta(\pi^+\pi^-) ~~~~~~{\rm or} ~~~~
[{\cal B}A_{CP}](K^+\pi^-) = -\frac{f_K}{f_\pi}[{\cal B}A_{CP}](\pi^+\pi^-)~,
\eeq
and
\bea
&&\Delta(K^{*+}\pi^-) =-\frac{f_{K^*}}{f_\rho}\Delta(\rho^+\pi^-) ~~~~{\rm or}~~~~
[{\cal B}A_{CP}](K^{*+}\pi^-) = -\frac{f_{K^*}}{f_\rho}[{\cal B}A_{CP}](\rho^+\pi^-)~,~~~~~\\
&&\Delta(\rho^-K^+)= -\frac{f_K}{f_\pi}\Delta(\rho^-\pi^+) ~~~~~~{\rm or}~~~~
[{\cal B}A_{CP}](\rho^-K^+) = -\frac{f_K}{f_\pi}[{\cal B}A_{CP}](\rho^-\pi^+)~.
\eea
Using branching ratios and asymmetries in Tables I and II, these three equalities read 
respectively in units of $10^{-6}$,
\bea
-1.90 \pm 0.23 & =  & -2.33 \pm 0.38~,\\
-1.54 \pm 0. 71 & = & -2.01 \pm 1.07~,\\
1.29 \pm 0. 54 & = & 1.27 \pm 1.03~.
\eea

The first test involves reasonably small errors and works well within $1\sigma$. 
It would have worked somewhat less well if penguin amplitudes were assumed to
factorize like tree amplitudes and to involve an SU(3) breaking factor $f_K/f_\pi$. 
The current experimental errors in $B\to VP$ decays are still very large  
and do not provide useful SU(3) tests.

An independent test for SU(3)  invariance of penguin amplitudes is provided by the ratio of
rates of penguin dominated $\Delta S=0$ and $\Delta S=1$ $B\to PP$ and $B\to VP$ 
decays,
\beq\label{P/Tratio}
\sqrt{\frac{{\cal B}(\bar K^0K^+)}{{\cal B}(K^0\pi^+)}} \simeq
\sqrt{\frac{{\cal B}(\bar K^{*0}K^+)}{{\cal B}(K^{*0}\pi^+)}} \simeq \tl~.
\eeq
Using ${\cal B}(\bar K^0K^+) = (1.36^{+0.29}_{-0.27})\times 10^{-6}$,
 ${\cal B}(\bar K^{*0}K^+) = (0.68 \pm 0.19)\times 10^{-6}$~\cite{HFAG}, and taking
 $\Delta S=1$ branching ratios in Table I, the above ratios of amplitudes are
 found to be $0.243 \pm 0.026$ and $0.262 \pm 0.038$, consistent with $\tl =0.232$ within 
 reasonable experimental errors. These errors must be reduced  somewhat in order to 
 distinguish between the assumed SU(3) invariant penguin amplitudes and SU(3) breaking 
 in these amplitudes given by a factor $f_K/f_\pi$.
 
\bigskip
{\bf VI. SU(3) RELATIONS BETWEEN TREE AND EWP AMPLITUDES}

\medskip
It has been noted that in the limit of flavor SU(3) symmetry approximate relations 
hold between the subdominant tree and electroweak penguin (EWP) amplitudes in 
$B\to K\pi$ transforming as $\Delta I=1$. Neglecting EWP operators ${\cal O}_7$ and 
${\cal O}_8$ with  tiny 
Wilson coefficients in the effective weak Hamiltonian, and using a quite precise relation for
Wilson coefficients (true within a few percent)~\cite{Buchalla:1995vs}, 
\beq\label{K}
{\cal K} \equiv \frac{c_9 + c_{10}}{c_1 + c_2} \approx \frac{c_9 - c_{10}}{c_1 - c_2}
\approx -0.0087~, 
\eeq
one obtains~\cite{Neubert:1998pt,Gronau:1998fn,Buras:1998rb},
\beq\label{EWPT1}
{\rm EWP}(B^+\to K^0\pi^+) + \s {\rm EWP}(B^+\to K^+\pi^0) = -(P_{EW} + P^C_{EW}) 
 =  \frac{3{\cal K}}{2}(T+C)~,
\eeq
and~\cite{Gronau:1998fn}
\beq\label{EWPT2}
{\rm EWP}(B^0\to K^+\pi^-) + {\rm EWP}(B^+ \to K^0\pi^+) = -P^C_{EW} 
 = \frac{3{\cal K}}{2}(C - E)~.
\eeq
These relations follow from corresponding properties of operators in the $\Delta S=1, \Delta I=1$
effective Hamiltonian behaving as $\overline{\bf 15}$ and ${\bf 6}$ under SU(3) 
transformation~\cite{Gronau:1998fn,Gronau:2000az},
\bea
\frac{1}{\lambda_t^{(s)}}{\cal H}^{(s)}_{\rm EWP}(\overline{\bf 15}) & = & 
-\frac{3{\cal K}}{2}\frac{1}{\lambda_u^{(s)}}{\cal H}_{\rm Tree}^{(s)}(\overline{\bf 15})~,\\
\frac{1}{\lambda_t^{(s)}}{\cal H}^{(s)}_{\rm EWP}(\bf 6) & = & 
\frac{3{\cal K}}{2}\frac{1}{\lambda_u^{(s)}}{\cal H}_{\rm Tree}^{(s)}({\bf 6})~,
\eea 
and imply relations similar to (\ref{EWPT1}) and (\ref{EWPT2}) in $B\to K^*\pi$ and  
$B\to \rho K$ decays.

It is convenient to study tree and EWP amplitudes for definite isospin states. 
We consider $\Delta I=1$ amplitudes for final states $f=K\pi, K^*\pi, \rho K$ with isospin 
$I=1/2, 3/2$, denoting corresponding tree and EWP contributions by ${\cal T}_I^f$, and 
${\cal E}_I^f$, respectively,
\beq
A^f_I = \lus{\cal T}^f_I + \lts{\cal E}^f_I~.
\eeq
Thus one has for instance,
\bea
6A^{K^*\pi}_{1/2} & = & \s A(K^{*+}\pi^0) + 4A(K^{*0}\pi^+) - 3\s A(K^{*0}\pi^0)
\nonumber\\
& \equiv &  6[\lus{\cal T}^{K^*\pi}_{1/2} + \lts{\cal E}^{K^*\pi}_{1/2}]~,\\
\label{A3/2}
3A^{K^*\pi}_{3/2} & = & A(K^{*+}\pi^-)+\s A(K^{*0}\pi^0) = A(K^{*0}\pi^+)+\s A(K^{*+}\pi^0)
\nonumber\\
& \equiv & 3[\lus{\cal T}^{K^*\pi}_{3/2}+\lts{\cal E}^{K^*\pi}_{3/2}]~.
\eea

We will apply flavor SU(3) to tree and EWP amplitudes 
in theses three classes of penguin-dominated decays. 
In the SU(3) limit both tree and EWP amplitudes may be 
expressed in terms of the same graphical contributions defined in 
Sections II~\cite{Gronau:1998fn} and III~\cite{Gronau:2000az},
\bea\label{TEKpi1}
6{\cal T}^{K\pi}_{1/2} & = & -T +2C + 3A~,~~~~~~~~~~~~~~~~
3{\cal T}^{K\pi}_{3/2}  = - T - C~,~\\
\label{TEKpi2}
6{\cal E}^{K\pi}_{1/2} & = & \frac{\cal K}{2}(-6T + 3C - 9E)~,~~~~~~~~~~
3{\cal E}^{K\pi}_{3/2} = \frac{3{\cal K}}{2}(T + C)~,~~~
\eea
\bea\label{TEK*pi1}
6{\cal T}^{K^*\pi}_{1/2} & = & -T_P +2C_V + 3A_P~,~~~~~~~~~~
3{\cal T}^{K^*\pi}_{3/2}  = - T_P - C_V~,\\
\label{TEK*pi2}
6{\cal E}^{K^*\pi}_{1/2} & = & \frac{\cal K}{2}(-6T_V + 3C_P - 9E_P)~,~~~~
3{\cal E}^{K^*\pi}_{3/2} = \frac{3{\cal K}}{2}(T_V + C_P)~,
\eea
\bea\label{TEKrho1}
6{\cal T}^{\rho K}_{1/2} & = & -T_V +2C_P + 3A_V~,~~~~~~~~~~~
3{\cal T}^{\rho K}_{3/2}  = - T_V - C_P~,\\
\label{TEKrho2}
6{\cal E}^{\rho K}_{1/2} & = & \frac{\cal K}{2}(-6T_P + 3C_V - 9E_V)~,~~~~~
3{\cal E}^{\rho K}_{3/2} = \frac{3{\cal K}}{2}(T_P + C_V)~.
\eea
We will make use of these expressions in the next section, assuming that the
annihilation-like amplitudes $A_{(P,V)}$ and $E_{(P,V)}$ are  negligible relative to the other 
amplitudes~\cite{Blok:1997yj,Bauer:2004ck}.

One may use the above relations to estimate the magnitudes of  EWP amplitudes
relative to those of dominant penguin amplitudes.
For instance, noting that the amplitude $T+C$ dominates  
$B^+\to \pi^+\pi^0$~\cite{Gronau:1994rj} with
${\cal B}(\pi^+\pi^0)=(5.59^{+0.41}_{-0.40})\times 10^{-6}$~\cite{HFAG} , 
while $P_{tc}$ (or actually $P_{tc} -P^C_{EW}/3$) governs $B^+\to K^0\pi^+$
in (\ref{0+}), Eq.~(\ref{EWPT1}) implies in the limit of flavor SU(3)~\cite{Gronau:1994bn}:
\beq
\frac{|P_{EW} + P^C_{EW}|}{|P_{tc}|} \simeq \frac {3|{\cal K}|}{\s}
\frac{|\lambda_t^{(s)}|}{|\lambda_u^{(d)}|}\sqrt{\frac{{\cal B}(\pi^+\pi^0)}{{\cal B}(K^0\pi^+)}}
= 0.10~,
\eeq
We use values of CKM factors $\lambda_q^{(q')}$ quoted in Refs.~\cite{PDG,Charles:2004jd}.
 We have not introduced a flavor SU(3) breaking factor $f_K/f_\pi$ in $T+C$ because
the  color-suppressed tree amplitude $C$, which is comparable in magnitude to $T$ 
with a large relative strong phase, cannot be assumed to  factorize.

A  ratio including CKM factors of an EWP amplitude and a corresponding tree amplitudes  
in $B^+\to K^+\pi^0$, which involve a common strong phase in the flavor SU(3) limit, 
is obtained directly from (\ref{EWPT1})~\cite{Neubert:1998pt,Gronau:1998fn}:
\beq
\frac{\lambda_t^{(s)}(P_{EW} + P^C_{EW})}{\lambda_u^{(s)}(T + C)} = 
-\frac{3{\cal K}}{2}\frac{\lambda_t^{(s)}}{\lambda_u^{(s)}} = -0.61e^{-i\gamma}~.
\eeq
Deviations from this pure SU(3) limit have been estimated and were found to be at 
most at a level of ten percent in the magnitude of the right-hand-side and a few degrees 
in its strong phase~\cite{Beneke:2006mk,Neubert:1998re}.

\bigskip
{\bf VII. CP ASYMMETRY SUM RULES IN $B\to K\pi, K^*\pi, \rho K$}

\medskip
A rather precise sum rule among the four CP rate differences in $B\to K\pi$ decays has been
proven in Ref.~\cite{Gronau:2005kz},
\beq
\Delta(K^+\pi^-) + \Delta(K^0\pi^+) - 2\Delta(K^+\pi^0) - 2\Delta(K^0\pi^0) \approx 0~.
\eeq
Using the approximate relation (\ref{ratios}) this implies a corresponding sum rule for
CP asymmetries given in Eq.~(\ref{SRasym}). 
This sum rule, which is based primarily on isospin
symmetry, provides a prediction for $A_{CP}(K^0\pi^0)$ in  terms of the other  more
precisely measured $B\to K\pi$ asymmetries. In this section we will study this sum rule and
similar ones for $B\to K^*\pi$ and $B\to \rho K$ decays, comparing the precision of the three
sum rules to one another.

We follow notations introduced in Section VI to denote by ${\cal T}^f_I$ and 
${\cal E}^f_I$ $\Delta I=1$ tree and EWP amplitudes for 
definite isospin states $I$, where $f=K\pi, K^*\pi, K\rho$. In addition, 
$\Delta I=0$ amplitudes multiplying  CKM factors $\lts$ ($P_{tc}$  and EWP) and 
$\lus$ ($P_{uc}$ and tree) will be denoted by ${\cal P}^f_{1/2}$ and $t^f_{1/2}$, respectively.
Subscripts on amplitudes will refer the charges of the two mesons  in the final state. 
Using isospin permits describing all twelve amplitudes for $B\to K\pi, K^*\pi, K\rho$, 
in the generic form,
\bea\label{f+-}
-A^f_{+-} & = & \lts\left ( {\cal P}^f_{1/2} - {\cal E}^f_{1/2} - {\cal E}^f_{3/2}\right) +
\lus\left(t^f_{1/2} - {\cal T}^f_{1/2} - {\cal T}^f_{3/2}\right) ~,\\
A^f_{0+} & = & \lts\left ( {\cal P}^f_{1/2} + {\cal E}^f_{1/2} + {\cal E}^f_{3/2}\right) +
\lus\left(t^f_{1/2} + {\cal T}^f_{1/2} + {\cal T}^f_{3/2}\right) ~,\\
-\s A^f_{+0} & = & \lts\left ( {\cal P}^f_{1/2} + {\cal E}^f_{1/2} - 2{\cal E}^f_{3/2}\right) +
\lus\left(t^f_{1/2} + {\cal T}^f_{1/2} - 2{\cal T}^f_{3/2}\right) ~,\\
\label{f00}
A^f_{00} & = & \lts\left ( {\cal P}^f_{1/2} - {\cal E}^f_{1/2} + 2{\cal E}^f_{3/2}\right) +
\lus\left(t^f_{1/2} - {\cal T}^f_{1/2} + 2{\cal T}^f_{3/2}\right) ~,
\eea 
where $f=K\pi, K^*\pi, K\rho$.

Defining CP rate differences for each of these twelve processes (common phase space 
factors for a given $f$ are omitted) ,
\beq
\Delta_{ij}^f\equiv |\bar A^f_{ij}|^2 - |A^f_{ij}|^2~,
\eeq
we consider the sums
\beq
\Delta(f) \equiv \Delta^f_{+-} + \Delta^f_{0+} - 2\Delta^f_{+0} - 2\Delta^f_{00}~.
\eeq
A generic amplitude of the form
\beq
A = \lts P + \lus T~,
\eeq
implies a CP rate difference (we omit a phase space factor)
\beq\label{Delgen}
\Delta \equiv |\bar A|^2 - |A|^2 = 4{\rm Im}(\lts\lusstar){\rm Im}(PT^*)~.
\eeq
Thus, using Eqs.~(\ref{f+-})-(\ref{f00}) one finds
\beq\label{SRf}
\Delta(f) = 24{\rm Im}[\lts\lusstar]{\rm Im}\left({\cal E}^f_{1/2}{\cal T}^{f*}_{3/2} + 
{\cal E}^f_{3/2}{\cal T}^{f*}_{1/2} - {\cal E}^f_{3/2}{\cal T}^{f*}_{3/2}\right )~.
\eeq

The dominant contributions to $\Delta(f)$, involving interference of the penguin 
amplitude (contained in ${\cal P}^f_{1/2}$) with tree amplitudes,  have cancelled.
This is the essence of the three asymmetry sum rules, shown in Ref.~~\cite{Gronau:2005kz} 
to follow from the isosinglet nature of the dominant penguin amplitude 
and an isospin quadrangle relation among the four $B\to f$ 
amplitudes~\cite{Lipkin:1991st,Gronau:1991dq},
\beq
-A^f_{+-} +A^f_{0+} +\s A^f_{+0} - \s A^f_{00} = 0~.
\eeq
A new $\Delta S= \Delta I=0$ operator in the effective Hamiltonian would not 
affect this argument as such an operator could be absorbed in the penguin operator.

The remaining terms in the sum rules (\ref{SRf}) involve second order interference 
terms of tree and EWP amplitudes. We will now estimate these remaining terms for
each of the three cases, $f=K\pi, K^*\pi, K \rho$, or suggest methods for measuring
these terms elsewhere. As we noted in Sec. IV, the potential
asymmetries in $B\to K^{*+}\pi^-$ and $B^+\to K^{*+}\pi^0$, or in $B^0\to K^+\rho^-$ 
and $B^+\to K^+\rho^0$, may be significantly larger than the $10\%$ asymmetry
measured in $B^0\to K^+\pi^-$. Thus naively one would expect the remaining terms in 
the $B\to K^*\pi$ and $B\to K\rho$ sum rules to be correspondingly larger than in the 
$B\to K\pi$ sum rule.

\bigskip
{\bf a. $B\to K\pi$}

\medskip
Inserting (\ref{TEKpi1}) and (\ref{TEKpi2}) into the sum rule (\ref{SRf}) for $f=K\pi$ and 
neglecting terms involving $A$ and $E$ one obtains
\beq\label{DelKpi}
\Delta(K\pi) = -12{\cal K}\,{\rm Im}[\lts\lusstar]{\rm Im}(CT^*)~.
\eeq
The sign of the remaining term in $\Delta(K\pi)$ is predicted to be negative because 
${\rm Im}[\lts\lusstar] = |V_{ts}||V_{cb}||V_{us}|\sin\gamma$ is positive while
${\cal K}$ and ${\rm Im}(CT^*)$ are negative. We have argued in Sec. II that the difference between 
$A_{CP}(K^+\pi^0)$ and $A_{CP}(K^+\pi^-)$ implies ${\rm Arg}(CT^*)<0$ or 
${\rm Im}(CT^*)<0$~\cite{Gronau:2006ha}.
 
The remaining term in $\Delta(K\pi)$ should be compared with the negative CP 
rate difference in $B^0\to K^+\pi^-$ which is dominated by an interference of 
$P$ and $T$,
\beq
\Delta(K^+\pi^-) \simeq 4{\rm Im}(\lts\lusstar){\rm Im}(PT^*)~.
\eeq
Using the numerical value of ${\cal K}$ in (\ref{K}) one has
\beq
\frac{\Delta(K\pi)}{\Delta(K^+\pi^-)} \simeq -3{\cal K} \frac{{\rm Im}(CT^*)}{{\rm Im}(PT^*)} =
0.026\frac{|C|}{|P|}\frac{\sin[{\rm Arg}(CT^*)]}{\sin[{\rm Arg}(PT^*)]}~.
\eeq
The ratio $|C|/|P|$, where the numerator and denominator do not include 
CKM factors, may be evaluated by assuming that $|C| \le |T|$ and taking our estimate 
$|T'|/|P'| \sim 0.1$ in Sec. IV. Using numerical values of CKM factors~\cite{PDG,Charles:2004jd}, 
this implies 
\beq
\frac{|C|}{|P|} \le \frac{|V_{tb}V_{ts}|}{|V_{ub}V_{us}|}\frac{|T'|}{|P'|} \sim 4.6~.
\eeq

One may further assume that the sine of the relative strong phase between $C$ and $T$ is not 
substantially larger than that of the relative strong phase between $P$ and $T$ (which we argued 
in Sec. IV to be around $30^\circ$ corresponding to $\sin[{\rm Arg}(PT^*)]=0.5$). This implies
\beq
0 \le \frac{\Delta(K\pi)}{\Delta(K^+\pi^-)} \le 0.12~.
\eeq
Although this positive upper bound is not rigid we consider it rather safe to conclude that the 
sum rule (\ref{SRasym}) for $B\to K\pi$ asymmetries  holds within two or perhaps even one percent,
\beq
-0.02~(-0.01)  < A_{CP}(K^+\pi^-) + A_{CP}(K^0\pi^+) - A_{CP}(K^+\pi^0) - A_{CP}(K^0\pi^0)  \le 0~.
\eeq
An important conclusion following from ${\rm Arg}(CT^*)<0$ is that the very small remaining term 
in the sum rule must be negative. Using three of the $B\to K\pi$ asymmetries in Table I leads to a prediction for  $A_{CP}(K^0\pi^0)$ which includes second order corrections in the asymmetry 
sum rule,
\beq
A_{CP}(K^0\pi^0) = -0.149 \pm 0.037 \pm 0.01~.
\eeq
The first error is purely experimental while the second one is due to a possible small violation
of the sum rule. 
We note that the current experimental errors in $A_{CP}(K^0\pi^+)$ and $A_{CP}(K^+\pi^0)$  
dominate the uncertainty in this prediction.

\bigskip
{\bf b. $B\to K^*\pi$ and $B\to K\rho$}

\medskip
Substituting (\ref{TEK*pi1}) and (\ref{TEK*pi2}) in the sum rule (\ref{SRf}) for $f=K^{*}\pi$ and 
neglecting small $A_P$ and $E_P$ terms one obtains,
\beq\label{DelK*pi}
\Delta(K^{*}\pi) = 6{\cal K}\,{\rm Im}[\lts\lusstar]{\rm Im}(T_VT^*_P + 2T_VC^*_V  + C_PC^*_V)~.
\eeq
As mentioned, expressions for $B\to K\rho$ amplitudes may be obtained from those for
$B\to K^*\pi$ by interchanging subscripts $V \leftrightarrow P$. This applies
also  to $\Delta(K\rho)$ which can be obtained from $\Delta(K^*\pi)$ through 
the same transformation:
\beq\label{DelKrho}
\Delta(K\rho) = 6{\cal K}\,{\rm Im}[\lts\lusstar]{\rm Im}(T_PT^*_V + 2T_PC^*_P  + C_VC^*_P)~.
\eeq

An interesting relation holds between the difference $\Delta(K\rho)-\Delta(K^*\pi)$ and a CP
rate difference for the amplitude $3A^{K^*\pi}_{3/2} \equiv  A(K^{*+}\pi^-)+\s A(K^{*0}\pi^0)$
already defined in (\ref{A3/2}),
\beq
\Delta\left((K^*\pi)_{3/2}\right) \equiv |3\bar A^{K^*\pi}_{3/2}|^2 - |3A^{K^*\pi}_{3/2}|^2~.
\eeq
Using Eqs.(\ref{A3/2}), (\ref{TEK*pi1}) and (\ref{TEK*pi2}),
\beq
3A^{K^*\pi}_{3/2} = -\lus(T_P + C_V) + \lts\frac{3{\cal K}}{2}(T_V + C_P)~,
\eeq
and applying (\ref{Delgen}) one has
\beq
\Delta\left((K^*\pi)_{3/2}\right)  = 6{\cal K}\,{\rm Im}[\lts\lusstar]
{\rm Im}[(T_P + C_V)(T^*_V + C^*_P)]~,
\eeq
which implyes
\beq\label{difDel}
\Delta(K\rho) - \Delta(K^*\pi) = 2\Delta\left((K^*\pi)_{3/2}\right)~.
\eeq
Note that, while certain individual CP rate asymmetries in $B\to K^*\pi$ and 
$B\to K\rho$ (involving interference of penguin and tree amplitudes) are expected to 
be large as discussed in Sec IV, the  asymmetry $\Delta((K^*\pi)_{3/2})$ involves 
interference of tree and EWP amplitudes and is considerably smaller.

The CP rate difference $\Delta((K^*\pi)_{3/2})$ can be measured in a Dalitz analysis of 
$B^0\to K^+\pi^-\pi^0$ and its charge-conjugate~\cite{Aubert:2007bs,Aubert:2008zu}. This 
analysis yields values for the magnitudes of $A(K^{*+}\pi^-), A(K^{*0}\pi^0)$, their relative
phase and their charge-conjugates which together fix 
$|3A^{K^*\pi}_{3/2} |$ and its charge conjugate thereby determining $\Delta((K^*\pi)_{3/2})$.
This quantity is related to a ratio of amplitudes, 
$R_{3/2} \equiv \bar A^{K^*\pi}_{3/2}/A^{K^*\pi}_{3/2}$, 
which is measurable in $B^0\to K^+\pi^-\pi^0$ and its charge conjugate:
\beq
\Delta\left((K^*\pi)_{3/2}\right) = |3A^{K^*\pi}_{3/2}|^2\left(|R_{3/2}|^2 - 1\right)~.
\eeq
The phase of $R_{3/2}$ has been proposed to give a new constraint on CKM 
parameters~\cite{Ciuchini:2006kv,Gronau:2006qn}. Its magnitude $|R_{3/2}|$ and  the 
magnitude $|3A^{K^*\pi}_{3/2}|$ determine $|\Delta\left((K^*\pi)_{3/2}\right)|$ which provides 
a good estimate for the combined violation of the $K^*\pi$ and $K\rho$ asymmetry sum rules.

Experimental errors using current data~\cite{Aubert:2008zu} are large and do not
permit a useful constraint on $\Delta((K^*\pi)_{3/2})$. 
The current error on $\Delta((K^*\pi)_{3/2})$ is comparable to the asymmetries measured
in $B\to K^*\pi$ decays. (This error decreases with increased statistics scaling roughly as 
$1/\sqrt{\rm Luminosity}$.)
A four-fold ambiguity in the solution
for $B\to K^*\pi$ amplitudes observed in Ref.~\cite{Aubert:2008zu}
may be resolved by requiring destructive interference between the two penguin-dominated
amplitudes $A(K^{*+}\pi^-)$ and $\s A(K^{*0}\pi^0)$ forming together the amplitude
$3A^{K^*\pi}_{3/2}$ which involves smaller tree and EWP contributions. This requirement applies
also to charge conjugate amplitudes, 
thus choosing solution I among the four solutions in Table V of Ref.~\cite{Aubert:2008zu}.

A somewhat less precise way for obtaining separate estimates for $\Delta(K^*\pi)$ and 
$\Delta(K\rho)$
requires using flavor SU(3) which relates $B\to K^*\pi$ and $B\to K\rho$ to $B\to \rho\pi$.
In the SU(3) symmetry limit, the right-hand-side of (\ref{DelK*pi}) may be expressed in terms of amplitudes for $B\to \rho\pi$ decays which are dominated by $T_{V,P}$ and $C_{V,P}$ 
(compare the first two equations with expressions in Table II)~\cite{Gronau:2000az}:
\bea\label{rho+pi-}
-A(\rho^+\pi^-) & \simeq & \lud T_P + \ltd P_P~,\\
-A(\rho^-\pi^+) & \simeq & \lud T_V + \ltd P_V~,\\
-2A(\rho^0\pi^0) & \simeq & \lud(C_V + C_P) - \ltd(P_V + P_P)~,\\
-\s A(\rho^+\pi^0) & \simeq & \lud(T_P + C_V) - \ltd(P_V - P_P)~,\\
\label{rho0pi0}
-\s A(\rho^0\pi^+) & \simeq & \lud(T_V + C_P) + \ltd(P_V - P_P)~.
\eea
We neglect tiny EWP contributions and very small terms $E_{V,P} + PA_{V,P}$, just as 
we have neglected $E+PA$ in $A(B^0\to \pi^+\pi^-)$ in Table II.

Working in the SU(3) symmetry limit, which is expected to introduce an uncertainty of 
$20-30\%$ in amplitudes, we will neglect contributions of penguin amplitudes 
in $B\to \rho\pi$ which will now be estimated to be at the same level. These 
contributions are measured by 
amplitudes  for $B\to K^*\bar K, \bar K^* K$.  Using the tree dominated branching ratio 
for $B^0\to \rho^+\pi^-$ in Table II and the penguin dominated branching ratio for 
$B^+\to \bar K^{*0} K^+$ quoted a line below Eq.~(\ref{P/Tratio}), one has
\beq
\frac{|\ltd P_P|}{| \lud T_P|} \simeq
\sqrt{\frac{{\cal B}(\bar K^{*0}K^+)}{r_\tau{\cal B}(\rho^+\pi^-)}} = 0.20 \pm 0.03~.
\eeq
This value is in agreement with the second range of values  in Eq.~(\ref{P'/T'}), obtained 
for the same ratio using somewhat different considerations including an SU(3) breaking 
factor $f_{K^*}/f_\rho$. 

In this approximation one finds
\bea
T_VT^*_P + 2T_VC^*_V  + C_PC^*_V & \simeq  & |\lud|^{-2}[2A(\rho^0\pi^+)A^*(\rho^+\pi^0) 
\nonumber\\
& + & \s A(\rho^-\pi^+)A^*(\rho^+\pi^0) - \s A(\rho^0\pi^+)A^*(\rho^+\pi^-)]~,
\eea
implying
\bea\label{DelK*pi2}
\Delta(K^{*}\pi) & \simeq & \frac{6{\cal K}\,{\rm Im}[\lts\lusstar]}{|\lud|^{2}}\,{\rm Im}
[2A(\rho^0\pi^+)A^*(\rho^+\pi^0) \nonumber\\
& + & \s A(\rho^-\pi^+)A^*(\rho^+\pi^0) - \s A(\rho^0\pi^+)A^*(\rho^+\pi^-)]~.
\eea
Charge conjugate modes may also be used on the right-hand-side to increase statistics.

The same quantity for $B\to K\rho$ may be obtained by interchanging subscripts 
$V\leftrightarrow P$ corresponding to interchanging the charges of $\rho$ and $\pi$
in $B\to \rho\pi$ amplitudes,
\bea\label{DelKrho2}
\Delta(K\rho) & \simeq & \frac{6{\cal K}\,{\rm Im}[\lts\lusstar]}{|\lud|^{2}}\,{\rm Im}
[-2A(\rho^0\pi^+)A^*(\rho^+\pi^0) \nonumber\\
& + & \s A(\rho^-\pi^+)A^*(\rho^+\pi^0) - \s A(\rho^0\pi^+)A^*(\rho^+\pi^-)]~.
\eea
Comparing this expression with  that of $\Delta(K^*\pi)$ in (\ref{DelK*pi2}), we note
that the only difference between the two expressions is the sign of the first term. 
Thus one may combine asymmetries as in (\ref{difDel}) to obtain:
\beq\label{Deldiff}
\Delta(K\rho) - \Delta(K^*\pi)  \simeq \frac{24{\cal K}\,{\rm Im}[\lts\lusstar]}{|\lud|^{2}}\,{\rm Im}
[A(\rho^+\pi^0)A^*(\rho^0\pi^+)]~.
\eeq

We now discuss a way by which the imaginary parts of products of $B\to\rho\pi$ amplitudes
in Eqs.~(\ref{DelK*pi2}), (\ref{DelKrho2}) and (\ref{Deldiff}) can be determined experimentally.  
Information about magnitudes and relative phases of the three $B^0$ decay amplitudes,
$A(\rho^+\pi^-), A(\rho^-\pi^+)$ and $A(\rho^0\pi^0)$,  
is obtained in a time-dependent Dalitz analysis of 
$B^0\to \pi^\pm\pi^\mp\pi^0$~\cite{Kusaka:2007dv,Aubert:2007jn},
$|A(\rho^0\pi^+)|$ is obtained in  a Dalitz analysis of 
$B^+\to \pi^+\pi^+\pi^-$~\cite{:2009az}, while $|A(\rho^+\pi^0)|$ is measured 
directly in this quasi two-body mode~\cite{Zhang:2004wza,Aubert:2007py}.
The five amplitudes obey an isospin pentagon relation~\cite{Lipkin:1991st,Gronau:1991dq},
\beq\label{pentagon}
A(\rho^+\pi^-) + A(\rho^-\pi^+) + 2A(\rho^0\pi^0) = \s A(\rho^+\pi^0) + \s A(\rho^0\pi^+) = D~,
\eeq
where $D$ is a diagonal of the pentagon describing an $I=2$ amplitude involving tree
contributions but no penguin amplitude. A similar relation holds for $\bar B$ decay amplitudes.

The magnitude $|D|$ can be determined by studying 
the three $B^0$ amplitudes in a Dalitz analysis of $B^0\to\pi^+\pi^-\pi^0$. 
The three magnitudes $|A(\rho^0\pi^+)|$, $|A(\rho^+\pi^0)|$ and $|D|$ fix the
triangle for these amplitudes up to a two-fold ambiguity corresponding to flipping the 
triangle around $D$. This gives a value for ${\rm Im}[A(\rho^+\pi^0)A^*(\rho^0\pi^+)]$. 
Information gained in $B^0\to\pi^+\pi^-\pi^0$ on magnitudes  of $A(\rho^\pm\pi^\mp)$ 
and their phases relative to $D$ permits a determination of 
${\rm Im}[A(\rho^-\pi^+)A^*(\rho^+\pi^0)]$ and ${\rm Im}[A(\rho^0\pi^+)A^*(\rho^+\pi^-)]$.

\begin{table}[h]
\caption{Branching fractions and CP asymmetries for $B\to \rho\pi$~\cite{HFAG}.
\label{tab:rhopi}}
\begin{center}
\begin{tabular}{c c c} \hline \hline
Mode & $\b$~$(10^{-6})$ & $A_{CP}$  \\ \hline\hline
$B^+ \to \rho^+\pi^0$ & $10.9^{+1.4}_{-1.5}$ & $0.02\pm 0.11$ \\
$B^+ \to \rho^0 \pi^+$ & $8.3^{+1.2}_{-1.3}$  & $0.18^{+0.09}_{-0.17}$\\
$B^0 \to \rho^0\pi^0$ & $2.0\pm 0.5$ & $-0.30\pm 0.38^{a}$ \\
\hline \hline
\end{tabular}
\end{center}
\leftline{$^a$ We take an average~\cite{HFAG} of Belle~\cite{Kusaka:2007dv} and 
Babar~\cite{Aubert:2007jn} results.}
\end{table}

In the approximation of neglecting penguin amplitudes in Eqs.~(\ref{rho+pi-})-(\ref{rho0pi0}),
(in which Eqs.~(\ref{DelK*pi2})-(\ref{Deldiff}) were written) or in the limit of
vanishing strong phases between tree and penguin amplitudes, all $B\to\rho\pi$  CP 
asymmetries vanish. Current experimental values of the five asymmetries
quoted in Table II and III are consistent with zero within experimental errors.
In this approximation the two pentagons for $B$ and $\bar B$ decays coincide, and 
Eq.~(\ref{pentagon}) turns into a single pentagon relation for square roots of decay rates.
A flat pentagon would correspond to
\beq
\sqrt{{\cal B}(\rho^+\pi^-)} + \sqrt{{\cal B}(\rho^-\pi^+)} + 2\sqrt{{\cal B}(\rho^0\pi^0)} = 
\s\sqrt{{\cal B}(\rho^+\pi^0)/r_{\tau}} + \s\sqrt{{\cal B}(\rho^0\pi^+)/r_{\tau}}~.
\eeq
Checking this possibility using branching ratios in Tables II and III, we find the following values 
(in units of $10^{-3}$) for the left and right-hand sides,
\beq\label{flatnumbers}
9.49 \pm 0.48 = 8.45 \pm 0.42~,
\eeq
which holds within $1.6\sigma$. In the limit of a flat pentagon the imaginary parts of all terms in 
(\ref{DelK*pi2}) and (\ref {DelKrho2}) vanish implying $\Delta(K^*\pi)=\Delta(K\rho)=0$.
Current agreement with a flat pentagon may indicate that violation of the $K^*\pi$ 
and $K\rho$ asymmetry sum rules are strongly suppressed.

Current information on $B\to\rho\pi$ amplitudes,  in particular that obtained from Dalitz 
analyses of $B^0\to\pi^+\pi^-\pi^0$ is not sufficiently precise for a useful quantitative study of 
$\Delta(K^*\pi)$ and $\Delta(K\rho)$. Errors are large in relative phases between amplitudes and 
in $B\to(\rho\pi)^0$ asymmetries.  
Belle measured a large central value for  $A_{CP}(\rho^+\pi^-)$ and a smaller value for 
$A_{CP}(\rho^-\pi^+)$~\cite{Kusaka:2007dv}, $A_{CP}(\rho^+\pi^-) = 0.21 \pm 0.08\pm 0.04, 
A_{CP}(\rho^-\pi^+)=0.08\pm 0.16 \pm 0.11$, whereas the situation in 
the Babar measurements was the opposite~\cite{Aubert:2007jn}, $A_{CP}(\rho^+\pi^-) = 
0.03 \pm 0.07\pm 0.04, A_{CP}(\rho^-\pi^+)=-0.37^{+0.16}_{-0.09}\pm 0.10$. 
Although the Belle and Babar results are statistically consistent with each other, this 
situation indicates large uncertainties. Using the Babar data and performing
a numerical study as described above, we obtain broad ranges for $\Delta(K^*\pi)$ and 
$\Delta(K\rho)$ which are consistent both with zero and with $\Delta(K^{*+}\pi^-)$ and 
$\Delta(K^+\rho^0)$, respectively.

A detailed study based on more precise data must also 
consider theoretical uncertainties in $\Delta(K^*\pi)$ and $\Delta(K\rho)$ caused by SU(3) 
breaking and by neglecting penguin amplitudes in 
$B\to\rho\pi$, which are given by decay amplitudes for $B\to K^*\bar K$ and $B\to \bar K^*K$. 
One may include these terms and smaller terms of the forms $E_{V,P} + PA_{V,P}$
by substituting into Eq.~(\ref{SRf})  the following expressions for the isospin 
amplitudes ${\cal T}_I^{K^*\pi}$ and ${\cal E}_I^{K^*\pi}$ in terms of $\Delta S=0$ 
amplitudes~\cite{Gronau:2000az}:
\bea
&& 6\lambda_u^{(d)}{\cal T}^{K^*\pi}_{1/2}=3 A(\rho^+\pi^-)- 
2\sqrt2 A(\rho^+\pi^0)-3 A(K^{*+}K^-)+
A(\bar K^{*0}K^+) + 2 A(K^{*+}\bar K^0),~~~~~\\
&&3\lambda_u^{(d)}{\cal T}^{K^*\pi}_{3/2} =  \sqrt2 A(\rho^+\pi^0) +
A(\bar K^{*0} K^+) - A(K^{*+} \bar K^0)\,,\\
&& \lambda_u^{(d)}{\cal E}_{1/2}^{K^*\pi} = \frac{{\cal K}}{4}[3A(\rho^-\pi^+) - 
 \sqrt2 A(\rho^0\pi^+) + 3 A(K^{*0}\bar K^0) + A(\bar K^{*0} K^+) -  A(K^{*+} \bar K^0)]\,,\\
&&\lambda_u^{(d)}  {\cal E}_{3/2}^{K^*\pi} = \frac{{\cal K}}{2}[-\sqrt2 A(\rho^0\pi^+) -  
A(K^{*+} \bar K^0) + A(\bar K^{*0} K^+)]~.
\eea
Corresponding expressions for the case $f=K\rho$ may be obtained from the above by
interchanging the charges of $\rho$ and $\pi$ and of $K^*$ and $K$.
Both magnitudes and relative phases of $B\to \rho\pi$ amplitudes may be 
measured as discussed above. This also applies to $B\to K^*\bar K$ and $B\to \bar K^*K$ amplitudes involving a common final $\bar KK\pi$ state.
However relative phases between the latter amplitudes and amplitudes for $B\to\rho\pi$ 
are unmeasurable.
 
\bigskip
{\bf VIII. CONCLUSION}

\medskip
A CP asymmetry of 10$\%$ has been measured in $B^0\to K^+\pi^-$. We have argued that 
asymmetries in $B^0\to K^{*+}\pi^-, K^+\rho^-$ and $B^+\to K^{*+}\pi^0, K^+\rho^0$ may be 
two or three times larger because of their larger ratios of tree and penguin amplitudes. 
A previously proven approximate isospin sum rule for CP asymmetries in $B\to K\pi$ 
decays was generalized to $B\to K^*\pi$ and $B\to K\rho$. The residues of the three sum 
rules consist of contributions from interference of tree and electroweak penguin amplitudes.
We have shown that the residue of the
$K\pi$ sum rule is negative at a level of one or at most two percent. Combining the  
$K^*\pi$ and $K\rho$ asymmetries into a single sum rule, we have shown that its residue 
is given by a CP asymmetry in $B\to (K^*\pi)_{I=3/2}$, which may be measured in a 
Dalitz analysis of $B^0\to K^+\pi^-\pi^0$. Using flavor SU(3), separate 
residues for the  $K^*\pi$ and $K\rho$ asymmetry sum rules may be obtained by studying 
$B\to\rho\pi$ amplitudes in $B$ decays into three pions. 

Some of the existing experimental results on $B\to K^*\pi$ and $B\to \rho K$ asymmetries
and on $B\to\rho\pi$ amplitudes do not use the full data samples accumulated for these 
modes at the two $e^+e^-$ $B$  factories, and do not study $K^*$ decays in all possible 
decay modes. We are awaiting more complete analyses. The latest results for 
$B^0\to K^+\pi^-\pi^0$ published by Belle~\cite{Chang:2004um} used a data sample from an
integrated luminosity of only $78$ fb$^{-1}$~\cite{Chang:2004um}. By now Belle has 
accumulated at least ten times more 
data for $B^0\to K^+\pi^-\pi^0$ which they should be encouraged to analyze. Measuring 
asymmetries at $B$ factories, at the LHCb detector and at a future Super Flavor Factory, in 
violation of these residues for $B\to K\pi$, $B\to K^*\pi$ or $B\to K\rho$ sum rules would provide 
evidence for  a new $\Delta S = \Delta I=1$ operator in the low energy effective Hamiltonian.
Such operators occur in a variety of extensions of the Standard Model~\cite{Browder:2008em}.

\bigskip
We thank Andrzej Buras and Andrew Wagner for questions which motivated this work and 
we are grateful to Tim  Gershon for a few useful comments. 
The work of JZ is supported in part by the European Commission RTN network, Contract 
No. MRTN-CT-2006-035482 (FLAVIAnet) and by the Slovenian Research Agency.


\end{document}